\documentclass[sigconf,authorversion,nonacm]{acmart}
\AtBeginDocument{%
  \providecommand\BibTeX{{%
    \normalfont B\kern-0.5em{\scshape i\kern-0.25em b}\kern-0.8em\TeX}}}

\begin{document}


\title{Technology-assisted Journal Writing for Improving Student Mental Wellbeing: Humanoid Robot vs. Voice Assistant}






\author{Batuhan Sayis}
\authornote{Undertook the work as a visiting researcher at the University of Cambridge.}
\affiliation{%
    \institution{Universitat Pompeu Fabra}
  \city{Barcelona}
  \country{Spain}}
\email{batuhan.sayis@upf.edu}

\author{Hatice Gunes}
\affiliation{%
    \institution{University of Cambridge}
  \city{Cambridge}
  \country{United Kingdom}}
\email{hatice.gunes@cl.cam.ac.uk}

\renewcommand{\shortauthors}{Batuhan Sayis and Hatice Gunes}

\begin{abstract}
Conversational agents have a potential in improving student mental wellbeing while assisting them in self-disclosure activities such as journalling. Their embodiment might have an effect on what students disclose, and how they disclose this, and student’s overall adherence to the disclosure activity. However, the effect of embodiment in the context of agent assisted journal writing has not been studied. Therefore, this study aims to investigate the viability of using social robots (SR) and voice assistants (VA) for eliciting rich disclosures in journal writing that contributes to mental health status improvement in students over time. Forty two undergraduate and graduate students participated in the study that assessed the mood changes (via Brief Mood Introspection Scale, BMIS), level of subjective self-disclosure (via Subjective Self-Disclosure Questionnaire, SSDQ), and perceptions toward the agents (via Robot Social Attributes Scale, RoSAS) with and without agent (SR or VA) assisted journal writing. Results suggest that only in robot condition there are mood improvements, higher levels of disclosure, and positive perceptions over time in technology-assisted journal writing. Our results suggest that robot assisted journal writing has some advantages over voice assistant one for eliciting rich disclosures that contributes to mental health status improvement in students over time.  

\end{abstract}

\ccsdesc[500]{Human-centered computing~Human computer interaction (HCI)}
\ccsdesc[500]{Computer systems organization~Robotics}

\keywords{conversational agents, embodiment, self-disclosure, journal writing, student mental health}


\maketitle

\section{Introduction}

\begin{figure*}[t]
\fbox{\includegraphics[width=175mm,height=40mm,scale=0.5]{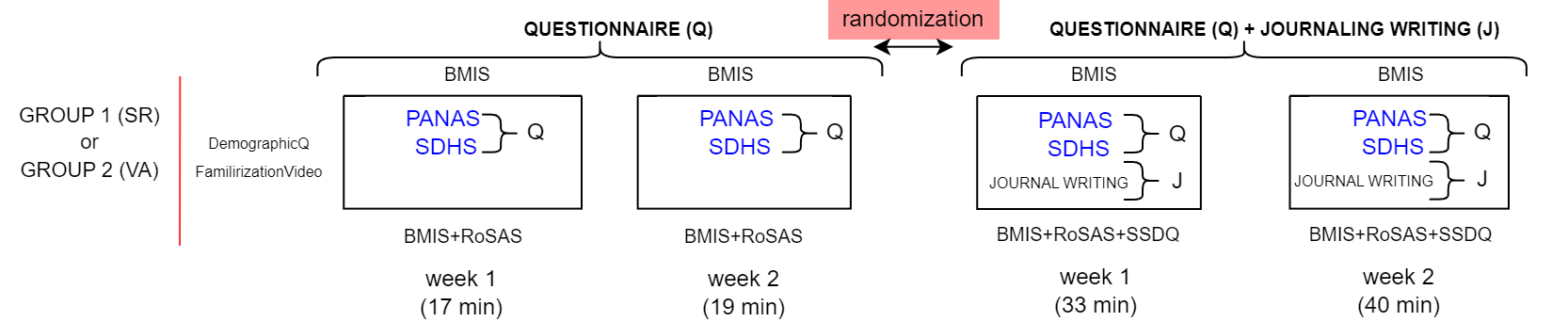}}
  \caption{Diagram of the experimental design. Abbreviations: GROUP1 (SR), Robot condition; GROUP2 (VA), Voice Assistant condition; BMIS, Brief Mood Introspection Scale; PANAS, Positive and Negative Affect Schedule; SDHS, The Short Depression-Happiness Scale; RoSAS, Robot Social Attributes Scale; SSDQ, Subjective Self-Disclosure Questionnaire
 }
\Description{This is a flow chart showing the information for the experimental design 2 by 2 by 2 mixed factorial design with between subject factor ‘’embodiment for assistance’’: robot condition, voice assistant); within subject factor ‘’activity’’: (questionnaire activity, questionnaire and journal writing activity); and within subject factor ‘’time’’: (week 1, week 2)}
\end{figure*}

High levels of stress combined with underdeveloped coping mechanisms and barriers in seeking support e.g. social stigma, shortage of mental health professionals, are contributing to a decline in university students’ mental health \cite{baik2019universities,jeong2020robotic}. Research has shown that self-disclosure in the form of journalling, specifically writing about one's experiences, feelings, and thoughts, can provide effective self regulation \cite{utley2011therapeutic,baikie2005emotional,hoyt2021emotional,sayis2024multimodal}. 
Conversational agents such as social robots, voice assistants and chatbots are gaining popularity due to their ability in encouraging honest and detailed self-disclosure about individual experiences and feelings \cite{lucas2017reporting,kawasaki2020assessing,van2014remote,ravichander2018empirical,abbasi2022can,duan2021self,bhakta2014sharing,spitale2023longitudinal}. Moreover, their embodiment has been reported to have an effect on what people disclose, and how they disclose it, and people’s adherence to the disclosure activities \cite{martelaro2016tell,laban2020let,laban2020tell,berry2005evaluating,powers2007comparing,lee2021exploring,van2014remote}. However, the effect of different technologies' embodiment has not been studied for writing based agent assisted journalling. Past works compared a control condition where there is no agent \cite{van2014remote,duan2021self} or with different versions of the same agent embodiment \cite{lee2021exploring,powers2007comparing}, 
where the agent is used either for mental health assessment \cite{kawasaki2020assessing} or mental health intervention \cite{li2023tell}. When agent assisted journalling was investigated as a possible mental health assessment strategy, the possible intervention effect of journalling tasks was not taken into account in the study designs (i.e., no control condition with self-assessment questionnaires) \cite{kawasaki2020assessing}. Therefore, our study aims to make a contribution in this direction by investigating the effect of embodiment in agent assisted journal writing, in a manner which allows the understanding of the intervention potential of technology-assisted journalling for student mental health. More specifically, this study 
explores the viability of using social robots (SR) and voice assistants (VA) for eliciting rich disclosures in journal writing that can contribute to mental health improvement in students over time. Accordingly, the research questions we investigate are: 
\textbf{(RQ1)} How does the (possible) improvement in mental health status of 
students differ between the VA and SR assisted journal writing? 
\textbf{(RQ2)} How do the students’ perceived self-disclosures change over time between the VA and SR assisted journal writing?
\textbf{(RQ3)} How do the students’ perception of the agent change over time between the VA and SR assisted journal writing?

\section{METHODS}

\subsection{Participants \& Study Design}

Sample size calculations and the study design were informed by similar research concerning each of our research questions. \cite{laban2020tell} answered a similar question to RQ2 \& RQ3 using a within subject design with 27 participants 
but was not focusing on the mental state improvement effect of agent assisted journal writing. 
\cite{li2023tell} answered a similar question to RQ1 using a within subject design with 42 participants 
but the writing activity used was not assisted by an agent. 
Considering all this, we designed a 2x2x2 mixed factorial design with between subject factor ‘’embodiment for assistance’’: (robot (SR), voice assistant (VA)); within subject factor ‘’activity’’: (questionnaire (Q), questionnaire + journal writing (Q+J)); and within subject factor ‘’time’’: (week 1, week 2) (Figure 1). Our study had 4 sessions in total and it was four weeks long (1 session per week). We recruited a total of 42 undergraduate and graduate students via flyers, posters in corridors, website, and newsletters on the campus of University of Cambridge. All students that enrolled in the study gave written informed consent before participation. While 40 of them completed all 4 sessions and received incentive of 30 GBP, the rest (2 in the VA condition) could not complete all the sessions and received 10 GBP for their participation. Data from 2 students were excluded from the analysis (from SR condition) due to missing self-report measures. The final sample consisted of 38 students (17 female, 19 male, 1 non-binary, 1 transgender male), 18 to 43 years old (M = 23.97 ± 5.41) (16: undergraduate students, 22: graduate students) (SR group(n): 19, VA group(n): 19).

\subsection{Agents Used}

As the social robot platform, we used the robot Pepper (SR) \cite{Aldebaran} which is 1.2 m tall 20\--DoF humanoid robot, equipped with multi-modal sensing capabilities and as the voice assistant we used ‘’Google Nest Mini’’ (VA) \cite{Nestmini} which has a compact footprint, and occupies minimal space since visual engagement is not its intended purpose. 
We chose SR and VA based on their popularity \cite{mishra2023exploration,Wardini_2023}, similar studies \cite{laban2023building,laban2020tell,forghani2023people} and their availability in the department.
VA served as a Bluetooth speaker, and to safeguard the privacy of students, both the Wi-Fi function and microphone were deactivated. They were clearly informed that the software for both agents is created by the experimenter, and it was emphasized that the Google Nest Mini does not have connection to Google web services.

\subsection{Agent assisted journal writing (J)}

The journal writing activity was adapted from \cite{meevissen2011become} and was generic in scope similarly to writing about one's weekly experiences, feelings, and thoughts. The activity consisted of 1 min of thinking and 10 mins of writing. 
As a journal writing tool a blank Microsoft Word page on a laptop without a Wi-Fi connection was used.
The task started by the following script being read by the agents: 
\textit{
‘The exercise you will do is to think about your schedule of the past week for one minute, and then write down your thoughts using the laptop provided.
These might include particular classes or meetings you attended, people you met, things you did etc. Go more deeply into the conversations, discussions, thoughts, or moods you may have had. Think of this as moving through your past week's schedule, day by day. Thus, you identify how this time period looks like for you. Now, please, start thinking about your past week's schedule, day by day. I will tell you when it is time to start writing down your thoughts.’} The agents did not intervene with the journal writing task, and they remained as a listener.
The social engagement of the agents was not adaptive based on the text entered into journals.

\subsection{Agent assisted questionnaire responding (Q)}

The agents administered the following questionnaires: Positive and Negative Affect Schedule (PANAS) \cite{watson1988development} and the Short Depression-Happiness Scale (SDHS) \cite{joseph2004rapid}. These questionnaires are acknowledged and accepted as standard instruments for documenting the immediate effects of an intervention, along with any results linked to positive psychological exercises, interventions, or activities. These tools have also been used in previous HRI research \cite{spitale2023robotic,axelsson2024oh,van2022tourists} 
which also informed us in adopting weekly sessions. Agents read out loud statements from PANAS (e.g., you have felt interested during the past week) and SDHS (e.g., you have felt dissatisfied with your life) while students verbalised their responses (i.e., for PANAS one option among “Very slightly or not at all”, “A little”, “Moderately”, ‘’Quite a bit’’, ‘’Extremely’’, and for SDHS one option among “Never”, ‘’Rarely’’, “Sometimes”, “Often”). The response options to each questionnaire were shown on the laptop screen, which served as a visual aid and relieved students from the need of memorizing these.

\subsection{Experimental Protocol}

The University of Cambridge Department of Computer Science and Technology Ethics Committee reviewed and approved the study protocol. 
Before coming to the first session, students were asked to fill out an online demographic questionnaire (DemographicQ).
Additionally, they were familiarised with the agent they were assigned to with a short video that they watched online. The study took place in a dedicated room within the department where each student interacted with one of the agents in a dyadic interaction setting monitored by the experimenter from a hidden control room. Each student was requested to sit on a chair that was placed 1.5m away from the agent positioned in front of the student. A laptop located in front of the student presented the tasks and activities that were run on PsychoPy (v2021.2.3). As presented in Figure 1, students completed a total of 4 sessions (2 with questionnaire activity (Q), 2 with questionnaire activity (Q) + journal writing (J)) across 4 weeks. During each session (see Figure 1)
students were first familiarised with the tasks after which the experimenter left the room to monitor the interaction from the control room. The same experimenter operated the agent (SR or VA as assigned to that student) via a Wizard of Oz (WoZ) setup from the control room. The agent assisted activities (J and Q) were pre-scripted and integrated into the WoZ setup. During each session, before and after agent assisted activities, on the laptop,
students filled in the Brief Mood Introspection Scale (BMIS, computes for sub-scores: Pleasant Unpleasant, Arousal-Calm, Positive-Tired and Negative-Relaxed Mood) \cite{mayer1988brief}. At the end of each session, on the laptop, students were asked to evaluate their perceptions toward the agents by filling the Robot Social Attributes Scale (RoSAS) \cite{carpinella2017robotic}. 
For the two sessions in which students were also assisted with J, after this activity, on the same laptop, students completed the Subjective Self-Disclosure Questionnaire (SSDQ) \cite{greenberg1996emotional} to measure their level of subjective self-disclosure. 
The assignment of the order of Q (week1 \& week2) and Q+J (week1 \& week2) was carried out randomly to counterbalance the order effect of the activities. 
   
\subsection{Data Analysis}

Students completed a total of 4 sessions (2 with Q, 2 with Q + J) across 4 weeks. In this work, we present an analysis of the SR and VA assisted activities and focus on self-report measures (BMIS, RoSAS \& SSDQ). Statistical analyses were undertaken using SPSS (v29). A series of repeated measures ANOVA with Greenhouse–Geisser correction were conducted for SR and VA separately, with conditions (Q vs. Q+J) and time (week1 (w1), week2 (w2)) set as independent variables, with each RoSAS sub-scale (competence, warmth, discomfort) as dependent variable. For the BMIS scores, similar analyses were conducted for SR and VA; and w1 and w2 separately with conditions (Q vs. Q+J) and time (preActivity (pre), postActivity (post)) set as independent variables, with BMIS sub-scale Negative-Calm as dependent variable. Adjustments for multiple comparisons were done with Bonferroni correction. On the other hand, as SSDQ was only answered in Q+J condition, for SR and VA separately, a series of paired sample T tests were conducted for each SSDQ subscale (personal, stressful, meaningful, expression of emotions, description of thoughts and/or feelings, awareness of thoughts and/or feelings).

\section{RESULTS}

Descriptive statistics are presented in Figures 2 \--- 4 (* represents p \begin{math} \leq \end{math} 0.05, ** represents p \begin{math} \leq \end{math} 0.01, *** represents p\begin{math} \leq \end{math} 0.001). 

Our first research question (RQ1) was to understand how the possible improvement in mental health status of students differ between the VA and SR assisted journal writing. For RQ1, when we checked the BMIS scores pre and post in each Q and Q+J, in each week, in the SR condition, results showed no significant main effect of the condition for the Negative-Calm mood levels. However, the result on the main effect was qualified by an interaction between activity and pre-post time, in w1. Checking the simple main effects revealed how Negative-Calm mood level of post Q+J is significantly lower than pre Q+J for w1. This was the case for w2 as well. However, when we ran the same analysis for the VA condition, this was not the case. In summary, it was possible to see significantly lower levels of negative affect (w1  \& w2) after each Q+J only in the SR condition (Fig. 2). 

For our second research question (RQ2),  we wanted to understand how the students’ perceived self-disclosure change over time between the VA and SR assisted journal writing. When we checked each SSDQ sub-scale levels regarding the differences between w1 and w2 (Fig. 3), in the SR condition we only see a significant difference in the sub-scale of awareness of thoughts and/or feelings. When we ran the same analysis for the VA condition, we did not see a significant difference between w1 and w2 in any of the SSDQ sub-scales. In summary, only in SR condition Q+J shows significant increase from w1 to w2 in perceived level of disclosure. 

For our third research question (RQ3), we wanted to understand how the students’ perception of the agent change over time between the VA and SR assisted journal writing. When we checked the RoSAS scores (Fig. 4) after each Q and Q+J, between w1 and w2, in the SR condition, results showed no significant main effect of the activity and no interaction effect between activity and time for all of the RoSAS subscale levels. This was the case for the VA condition as well. However, checking the simple main effects reveals that only in SR condition there is a significant decrease from w1 to w2 in discomfort.  

\begin{figure}[h]
  \centering
  \includegraphics[width=\linewidth]{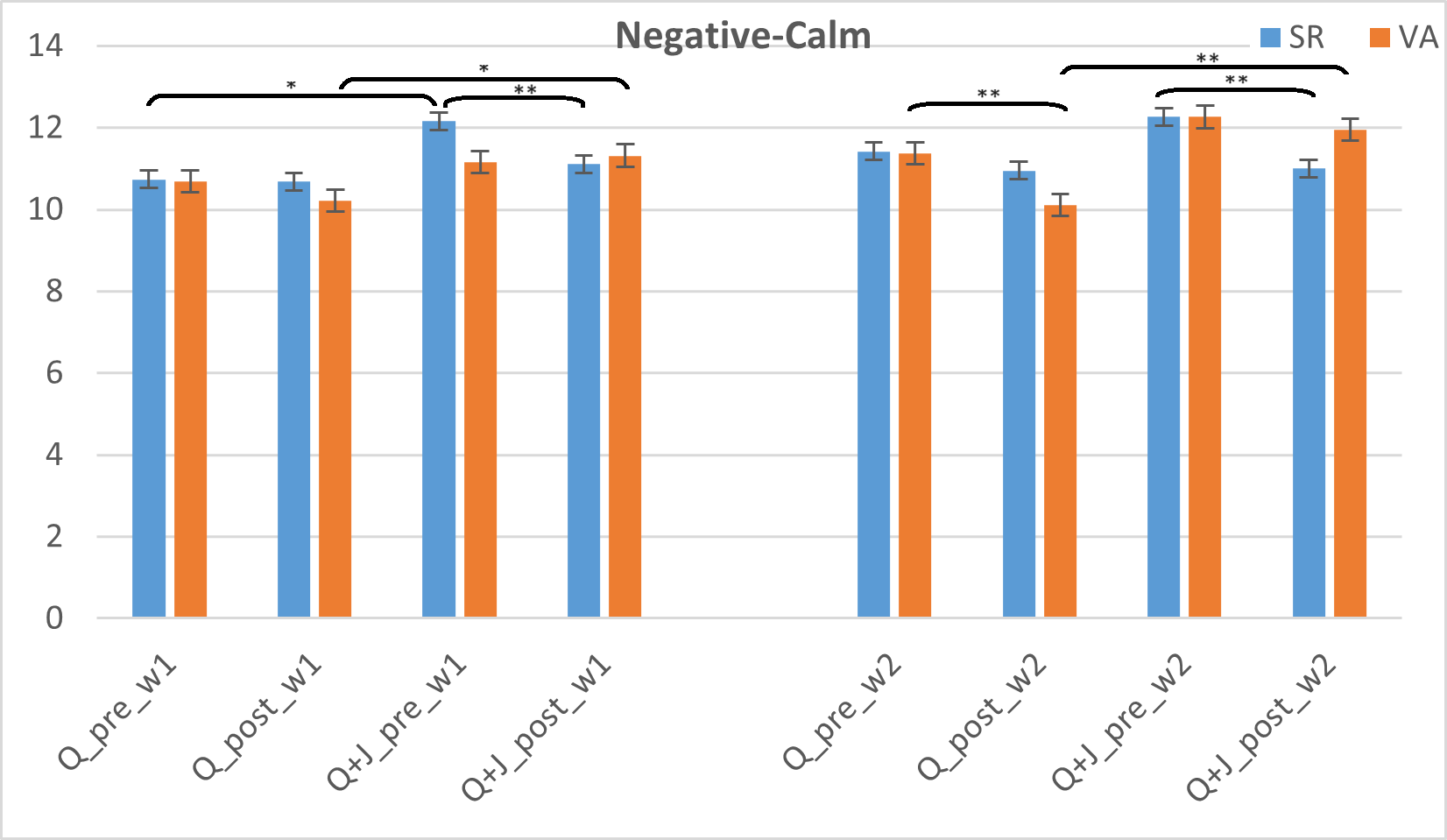}
\caption{Mean BMIS Negative-Calm sub-scale levels pre \&  post, in Q \&  Q+J, in week1 (w1) \& week2 (w2). Lower values indicate less negative mood.}
\Description{This is a bar graph showing the Mean Brief Mood Introspection Negative-Calm sub-scale levels pre activity and post activity, in questionnaire activity and questionnaire and journal writing activity together, in week1 and week2. Only in robot condition questionnaire and journal writing activity together in comparison to questionnaire activity alone from pre to post activity leads to significantly lower levels of negative affect both in week1 and week2. Bars with blue colour shows results from robot condition and the orange colour shows results from voice assistant condition.}
\end{figure}

\begin{figure}[h]
  \centering
  \includegraphics[width=\linewidth]{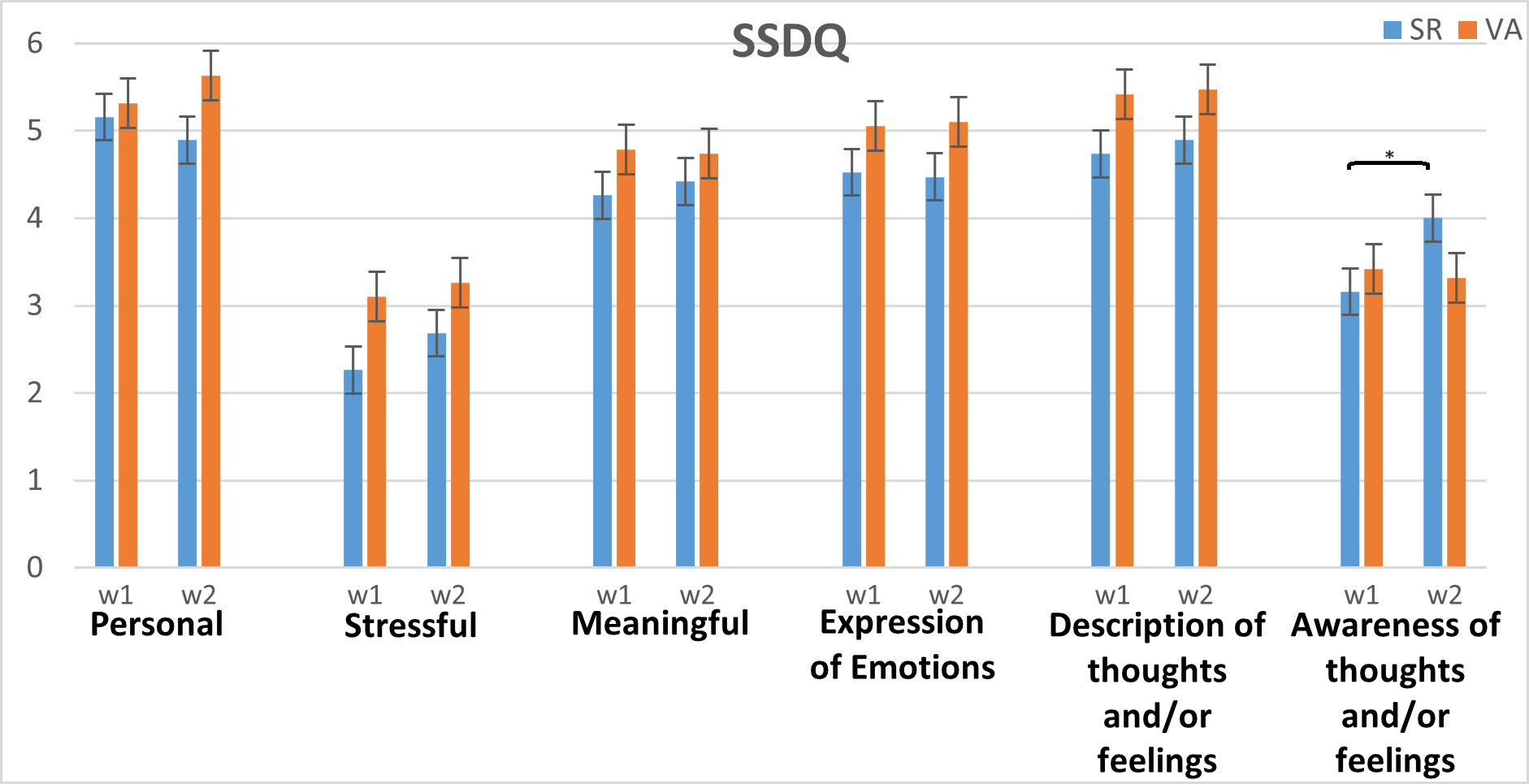}
    \caption{Mean SSDQ sub-scale levels, after Q+J, in week1 (w1) \& week (w2). Only in SR, Q+J shows significant increase from w1 to w2 in the awareness of thoughts and/or feelings.}
\Description{   This is a bar graph showing the Mean Subjective Self-Disclosure Questionnaire sub-scale levels after questionnaire and journal writing activity together, in week1 and week2. Only in robot condition questionnaire and journal writing activity together shows significant increase from week1 to week2 in the sub-scale of awareness of thoughts and/or feelings. Bars with blue colour shows results from robot condition and the orange colour shows results from voice assistant condition.}

\end{figure}

\begin{figure}[h]
  \centering
  \includegraphics[width=\linewidth]{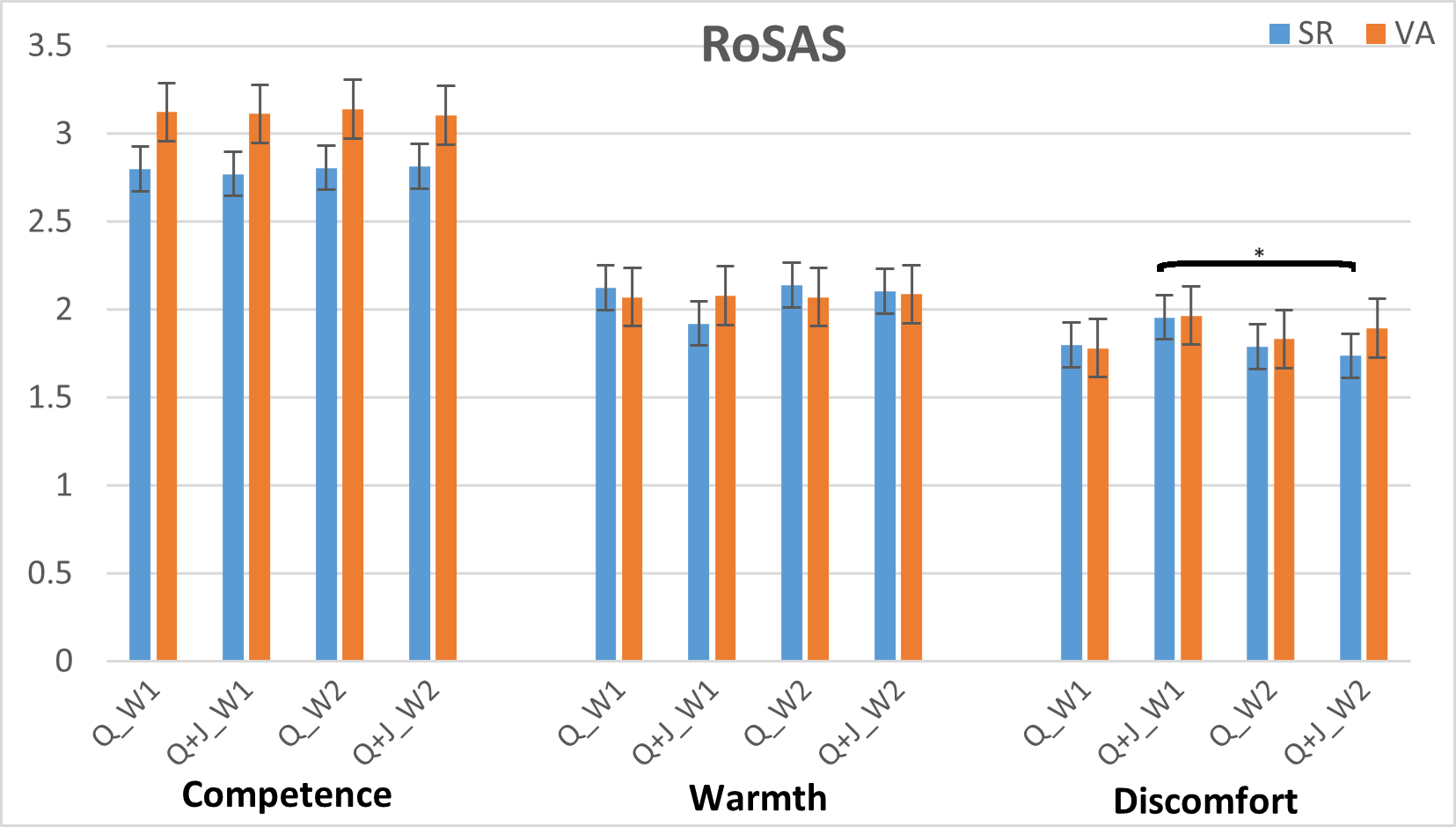}
      \caption{Mean RoSAS sub-scale levels, after Q \& Q+J, in week (w1) \& week (w2). Only in SR, Q+J shows significant decrease from w1 to w2 in the sub-scale of discomfort.}
\Description{This is a bar graph showing Mean Robot Social Attributes Scale sub-scale levels after questionnaire activity and questionnaire and journal writing activity together, in week1 and week2. Only in robot condition questionnaire and journal writing activity together shows significant decrease from week1 to week2 in the sub-scale of discomfort. Bars with blue colour shows results from robot condition and the orange colour shows voice assistant condition.}
\end{figure}


\section{DISCUSSION}

Results for RQ1 suggest that only in SR condition Q+J in comparison to Q from pre to post activity leads to significantly lower levels of negative affect (w1 \& w2). Moreover, results  for RQ2 suggest that only in SR condition, Q+J shows significant increase from w1 to w2 in level of perceived disclosure specifically in awareness of thoughts and/or feelings. Additionally, results for RQ3 suggest that only in SR condition, Q+J shows significant decrease from w1 to w2 in level of discomfort. These results together suggest that robot assisted journal writing has some advantage over VA assisted journal writing for eliciting rich disclosures that contributes to mental health status improvement in students over time. 

In relation to effects in mood improvements (RQ1) in Q+J activity in comparison to Q activity, only in SR condition results are consistent with previous literature where journalling significantly improved mood pre to post activity \cite{meevissen2011become,heekerens2021inducing,utley2011therapeutic}. With the assumption that journalling should improve mood whether it is with a robot or VA, we were expecting both conditions would be effective so that we could compare how this effect differs between Robot and VA. The explanation for this might be related to the findings of \cite{li2023tell} where speaking into an audio recorder (solo speaking) in the form of journalling decreased momentary mood and increased momentary stress. Similarly to the discomfort associated with speaking without an audience, in our study, students, while going through VA assisted journal writing might have had a similar feeling \--- i.e., the physical embodiment of VA resembles talking to an audio recorder, that resulted in no mood improvements with Q+J in VA. 

In relation to the effects in level of disclosures (RQ2), only in SR condition, and only in one SSDQ subscale, Q+J shows significant increase from w1 to w2, specifically in awareness of thoughts and/or feelings. Similar results have been reported by \cite{laban2020tell}, but disclosing in a speech format not in a writing format. People perceived that they disclosed less information to a VA than to a humanoid robot. \cite{laban2020tell} discussed that limits in conveying an involvement through body cues in an interaction could explain the low level of disclosure. This interpretation might apply to our setting as well.

In relation to the effects in the students’ perception of the agents (RQ3) only in SR condition, and only in  discomfort subscale, Q+J shows significant decrease from w1 to w2. Research has suggested that integrating journal writing into counseling sessions serves as an innovative method to involve clients in a therapeutic activity \cite{utley2011therapeutic}. This practice can foster increased self-awareness and personal growth, not only within the session but also between sessions. Furthermore, the activity may be associated with cultivating a sense of comfort in self-intimacy, laying the groundwork for establishing deeper connections with a therapist. In a similar vein, we hypothesise that SR condition might have brought students to a more comfortable state \---i.e., they were more aware of their thoughts and/or feelings. This interpretation is consistent with the findings from BMIS Negative-Calm sub-scale where in SR condition, both in w1 \& w2, with Q+J there was a significant difference between pre and post activity which might signal a consistent sense of comfort.

\section{LIMITATIONS \& FUTURE WORK}

While our study results show promise, it is important to note several limitations. First, VA condition results were unexpected as Q+J activity should have been effective in VA assistance as well. To investigate this further, Q and J should also be evaluated separately in future studies. Second, the Independent Samples t tests should also be included in the analyses in order to determine whether there is statistical evidence that the associated population means for SR and VA conditions are significantly different. 
Third, self-report measures can be influenced by biases such as honesty, introspective ability, and the interpretation of questions. Analysis from other measures such as journal entries (content analysis) or sensor data (e.g., physiological and/or visual cues) should also be considered for a more thorough understanding. 
Fourth, further research should explore the generalizability of the comparison between VA and SR using different social robots.
Finally, even though the study design (including instruments and protocol) can be modified for real-world scenarios (e.g., home settings or school counseling services), the current investigation was carried out in a laboratory setting. Although this setting reduces external validity it facilitated the assessment of the activities. Future research in this area should consider addressing the aforementioned limitations.

\vspace{-2mm}
\begin{acks}

B. Sayis is supported by European Union-NextGenerationEU, Ministry of Universities and Recovery, Transformation and Resilience Plan, through a call from Universitat Pompeu Fabra (Barcelona) and by TIDE through the Universitat Pompeu Fabra Initiatives (Planetary Wellbeing: PLAWB00322). H. Gunes is supported by the EPSRC/UKRI under grant ref. EP/R030782/1 (ARoEQ).

\noindent \textbf{Open Access:} For open access purposes, the authors have applied a Creative Commons Attribution (CC BY) licence to any Author Accepted Manuscript version arising.
\end{acks}

\bibliographystyle{ACM-Reference-Format}
\balance
\bibliography{sample-base}

\appendix

\end{document}